\documentclass[aps,superscriptaddress,twocolumn]{revtex4-1}
\usepackage{amsfonts}
\usepackage{subfigure}
\usepackage{txfonts}
\usepackage{amssymb}
\usepackage{amsbsy} 
\usepackage{epsfig}
\usepackage{color}

\def\be{\begin{equation}} \def\ee{\end{equation}}
\def\bea{\begin{eqnarray}} \def\eea{\end{eqnarray}}

\def\bq{{\bf q}}
\def\bQ{{\bf Q}}
\def\bk{{\bf k}}

\def\bK{{\bf K}}
\def\be{{\bf e}}

\def\bA{{\bf A}}
\def\bn{{\bf n}}
\def\be{{\bf e}}

\begin{document}

\title{Floquet multi-Weyl points in crossing-nodal-line semimetals}

\author{Zhongbo Yan}
 \affiliation{ Institute for
Advanced Study, Tsinghua University, Beijing,  China, 100084}

\author{Zhong Wang}
\affiliation{ Institute for
Advanced Study, Tsinghua University, Beijing,  China, 100084}

\affiliation{Collaborative Innovation Center of Quantum Matter, Beijing 100871, China }

\date{\today}

\begin{abstract}

Weyl points with monopole charge $\pm 1$ have been extensively studied, however, real materials of multi-Weyl points, whose monopole charges are higher than $1$, have yet to be found. In this Rapid Communication, we show that nodal-line semimetals with nontrivial line connectivity provide natural platforms for realizing Floquet multi-Weyl points. In particular, we show that driving crossing nodal lines by circularly polarized light generates double-Weyl points. Furthermore, we show that monopole combination and annihilation can be observed in crossing-nodal-line semimetals and nodal-chain semimetals.
These proposals can be experimentally verified in pump-probe angle-resolved photoemission spectroscopy.
\end{abstract}

\maketitle

Stimulated by extensive studies on topological insulators \cite{hasan2010,qi2011,Chiu2016rmp}, it has now been realized that many metals also have topological characterizations\cite{Chiu2016rmp,Armitage2017,Zhao2013classification}. In these topological semimetals, the valence band and conduction band touch at certain $\bk$-space manifolds. When the band-touching manifolds consist of isolated points, the materials are nodal-point semimetals, Dirac semimetals\cite{liu2014discovery,neupane2014,Borisenko2014,xu2015observation, wang2012dirac,young2012dirac,wang2013three,Sekine2014,Zhang2015detection,Yang2014dirac} and Weyl semimetals(WSMs)\cite{wan2011,murakami2007phase,nielsen1983adler,volovik2003,yang2011,
burkov2011,Son2013chiral, wang2013a, lu2013weyl,Hosur2013,Lu2014review, weng2015,Huang2015TaAs,Xu2015science,lv2015,Huang2015,
Ghimire,Shekhar,Xu2015NbAs,lu2015,Zhou-plasmon,Bi2015, Lu2016review,Yan2016review} being the most well-known examples; when the band-touching manifolds are one-dimensional lines, the systems are nodal-line semimetals(NLSMs)\cite{Burkov2011nodal,
Carter2012,
Phillips2014tunable,chen2015topological,
Chiu2014,Mullen2015,Bian2015nodal,
Xie2015ring,Chen2015line,Fang2015nodal,Bian2015TlTaSe,Chan2015ca3p2,Zeng2015nodal,Weng2015nodal,Kim2015,Yu2015,
Gan2016XB6, Kawakami2016line, Bzdusek2016,Hirayama2017nodal,Zhao2016line,
Liang2016line, Li2016line, Sung2016line,Rhim2015Landau,
schoop2016line,Neupane2016,Singha2016, Hu2016line, Wu2016PtSn4,
Rhim2016nodal,Yan2016nodal, Mikitik2016nodal, Roy2016line, Sur2016line, Liu2016line,Li2017line,fang2017}.

In WSMs, the Weyl points are the sources or sinks of Berry magnetic field,
namely, they are the Berry monopole charges. The total number of monopole
charge in the Brillouin zone must be zero, which has been formulated decades
ago as a no-go theorem by Nielsen and Ninomiya\cite{NIELSEN1981}. Usually,
a Weyl point has a linear dispersion in all three spatial directions, with
a low-energy Hamiltonian $H(\bq)=\sum_{i,j=x,y,z}v_{ij}q_i\tau_j$, where
$\tau_{x,y,z}$ are the Pauli matrices. The monopole charge is just
$C={\rm sgn}[\det(v_{ij})]=\pm 1$. Interestingly, multi-Weyl points with
monopole charge higher than one are also possible. The simplest cases are
the double-Weyl points with $C=\pm2$\cite{Xu2011,Fang2012multiweyl,Huang2016double}, which have novel physical consequences\cite{Lai2015doubleweyl,Jian2015doubleweyl,
Chen2016doubleweyl,Ahn2016multiweyla,Ahn2016multiweylb,Gupta2017floquetweyl}.
So far, the double-Weyl points have not been experimentally realized in
solid-state materials. Considering the widespread interests in WSMs,
it is highly interesting to find material realizations of multi-Weyl points.
The combination of several Weyl points into a multi-Weyl point, and
the annihilation of several Weyl points are even more interesting to
investigate, nevertheless, it is  challenging to do so because of the limited tunability in the samples.

Over the past few years, periodic driving has been used as a powerful
method to alter the topology of static systems, and more remarkably, to create new
topological phases without analog in static systems\cite{lindner2011floquet,Kitagawa2011,Oka2009,Inoue2010,Gu2011,
Kitagawa2010a,Kitagawa2010b,Lindner2013,Jiang2011,rudner2013anomalous,Dahlhaus2011,Gomez2013,
Zhou2011Optical,Delplace2013,wang2013observation,Perez2014graphene,
mahmood2016selective, Giovannini2016floquet,Chen2016floquet, Qu2016graphene, Liu2016graphene,bi2016}.
Recently, there are a few
theoretical proposals for Floquet topological semimetals\cite{wang2014floquet,Narayan2015dirac,
Bomantara2016Harper,Zou2016floquet,Wang2016floquet,Chan2016hall,Yan2016tunable, Chan2016nodal,
Narayan2016nodal,
Hubener2016,Zhang2016floquet,Hashimoto2016floquet}, in particular,
it has been suggested that under a circularly polarized light (CPL), NLSMs will be driven to Floquet WSMs with
highly tunable Weyl points~\cite{Yan2016tunable, Chan2016nodal,
Narayan2016nodal,Taguchi2016nodal}. In these studies, only the simplest nodal lines are considered.
The present work is stimulated by recent proposals of novel nodal lines with
nontrivial connectivity, including crossing nodal lines\cite{Kim2015,Yu2015,
Du2016CaTe,Kobayashi2017} (probably the most interesting ones are the nodal
chains\cite{Bzdusek2016,Yu2017chain,wang2017chain}), nodal links\cite{hopflink,nodal-link,weyl-link,Ezawa2017},
and nodal knots\cite{bi2017knot}. In this Rapid Communication, we show that crossing nodal
lines (including nodal chains) are natural platforms for the realizations of Floquet multi-Weyl points
and the combinations (annihilations) of Weyl points. In particular, a two-nodal-line crossing point
can be driven to a double-Weyl point, and tuning the direction of incident lasers can induce monopole
combination transitions. Considering the abundant material candidates for crossing nodal
lines\cite{Zeng2015nodal,Weng2015nodal,Kim2015,Yu2015, Du2016CaTe, Gan2016XB6,
Kawakami2016line,Bzdusek2016}, we believe that this proposal can be experimentally verified in the near future.

\begin{figure}
\subfigure{\includegraphics[width=4.25cm, height=4cm]{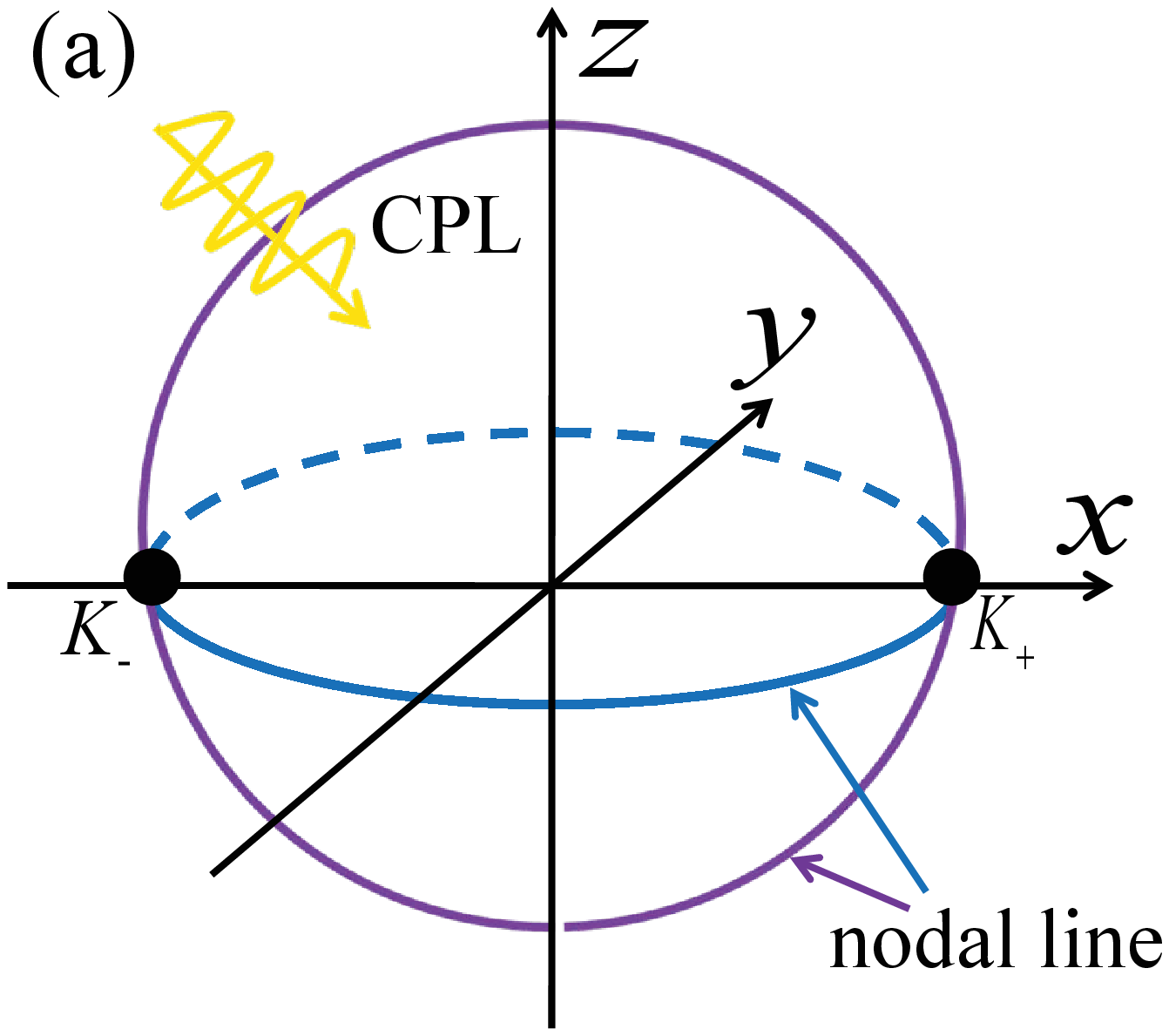}}
\subfigure{\includegraphics[width=4.25cm, height=4cm]{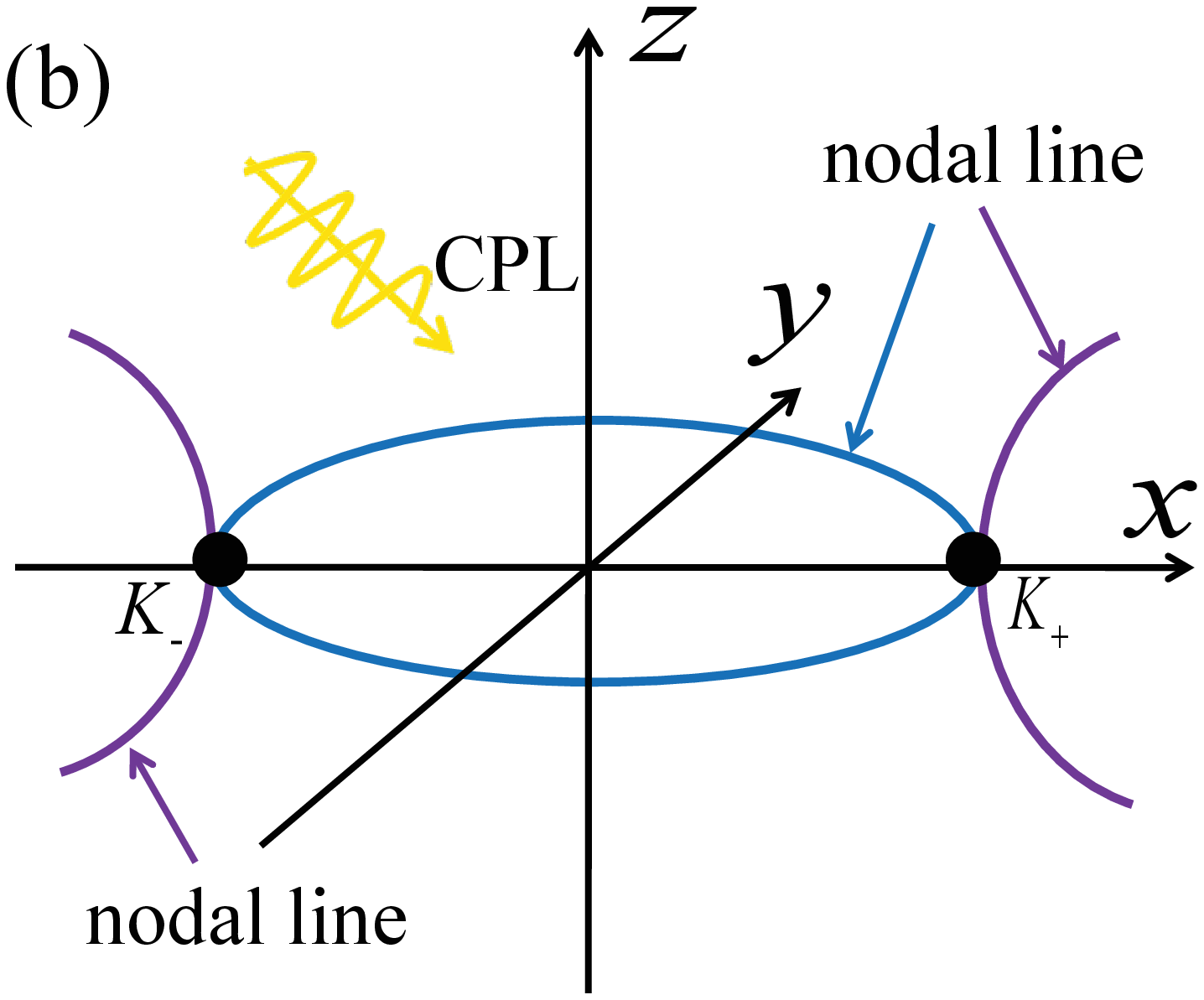}}
\caption{ Two types of nodal-line crossing.   (a) Type-I crossing: two
nodal lines are located on the same side of the tangential plane (the $k_{x}=\pm |\bK_{+}|$ planes)
near their crossing points.
(b) Type-II crossing: two
nodal lines are located on the opposite side of the tangential plane near their crossing
points.
}  \label{sketch}
\end{figure}

{\it Double-Weyl points from Type-I crossing.---}
For simplicity, we focus on NLSMs with negligible spin-orbit
coupling\cite{Kim2015,Yu2015}. We distinguish two types of
nodal line crossing, illustrated in Fig.\ref{sketch}(a) and (b),
as type-I and type-II crossing, respectively. The type-II crossing
is the basic building block of nodal chains. In this section, we focus on the type-I crossing.
Our starting point is the following Bloch Hamiltonian ($\hbar=c=k_{B}=1$)
\begin{eqnarray}
H(\bk)=(m-Bk^{2})\tau_{x}+\lambda k_{y}k_{z}\tau_{z} + \epsilon_{0}(\bk)\tau_0,\label{nlsm}
\end{eqnarray}
where $\tau_{x,y,z}$ are Pauli matrices in orbital space and
$\tau_0$ is the identity matrix, $m$ is a positive constant with
the dimension of energy, $B$ and $\lambda$ are positive constants
with the dimension of inverse energy, and  $k^{2}=k_{x}^{2}+k_{y}^{2}+k_{z}^{2}$.
As the diagonal term $\epsilon_0(\bk)$ does not affect the main physics,
we will neglect it hereafter.
The energy spectra of this Hamiltonian read
\begin{eqnarray}
E_{\pm,\bk}=\pm\sqrt{(m-Bk^{2})^{2}+\lambda^{2}k_{y}^{2}k_{z}^{2}}.
\end{eqnarray}
It is readily found that there are two nodal lines, one is located
in the $k_z=0$ plane and determined by the equation
$k_{x}^{2}+k_{y}^{2}=m/B$, while the other one is located in the $k_y=0$ plane and
determined by the equation $k_{x}^{2}+k_{z}^{2}=m/B$. The two
nodal lines cross at $\bK_{\pm}=\pm(\sqrt{m/B},0,0)$, which gives the type-I
crossing illustrated in Fig.\ref{sketch}(a).
The crossing points are protected by the mirror symmetry:
$\mathcal{M}_{z}H(k_{x},k_{y},k_{z})\mathcal{M}_{z}^{-1}=H(k_{x},k_{y},-k_{z})$
and $\mathcal{M}_{y}H(k_{x},k_{y},k_{z})\mathcal{M}_{y}^{-1}=H(k_{x},-k_{y},k_{z})$
with $\mathcal{M}_{z}= \mathcal{M}_{y}= \tau_{x}$.

We study the effects of a CPL. Let us consider a CPL incident
in the direction $\bn=(\cos\phi\sin\theta,\sin\phi\sin\theta,\cos\theta)$, where
$\theta$ and $\phi$ is the polar and azimuthal angle in the spherical coordinate system, respectively.
The  vector potential of the light is $\bA(t)=A_{0}(\cos(\omega t)\be_{1}
+\eta\sin(\omega t)\be_{2})$, with $\eta=\pm1$ corresponding to the right-handed
and left-handed CPL, respectively. Here, $\be_{1}=(\sin\phi,-\cos\phi,0)$
and $\be_{2}=(\cos\phi\cos\theta,\sin\phi\cos\theta,-\sin\theta)$ are two unit
vectors perpendicular to $\bn$, satisfying $\be_{1}\cdot\be_{2}=0$.

\begin{figure}
\subfigure{\includegraphics[width=6cm, height=4cm]{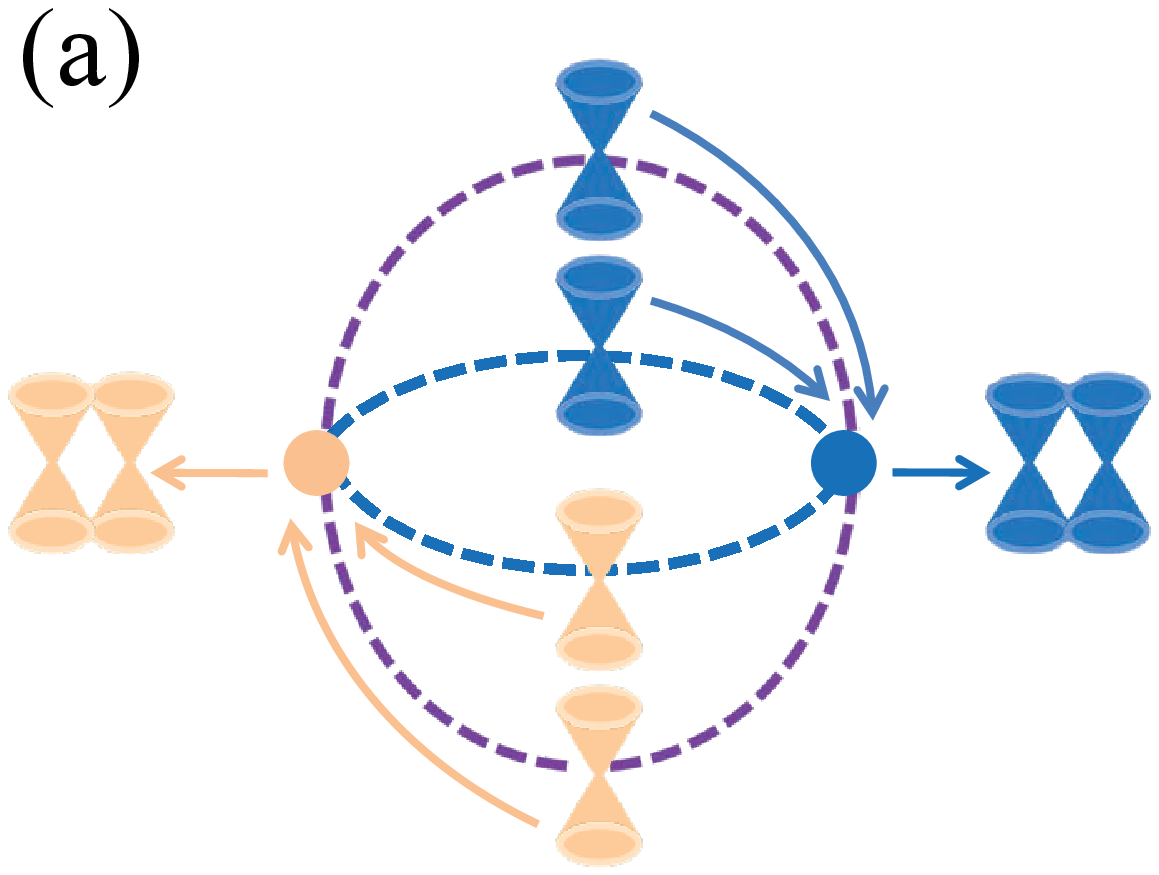}}
\subfigure{\includegraphics[width=6cm, height=4cm]{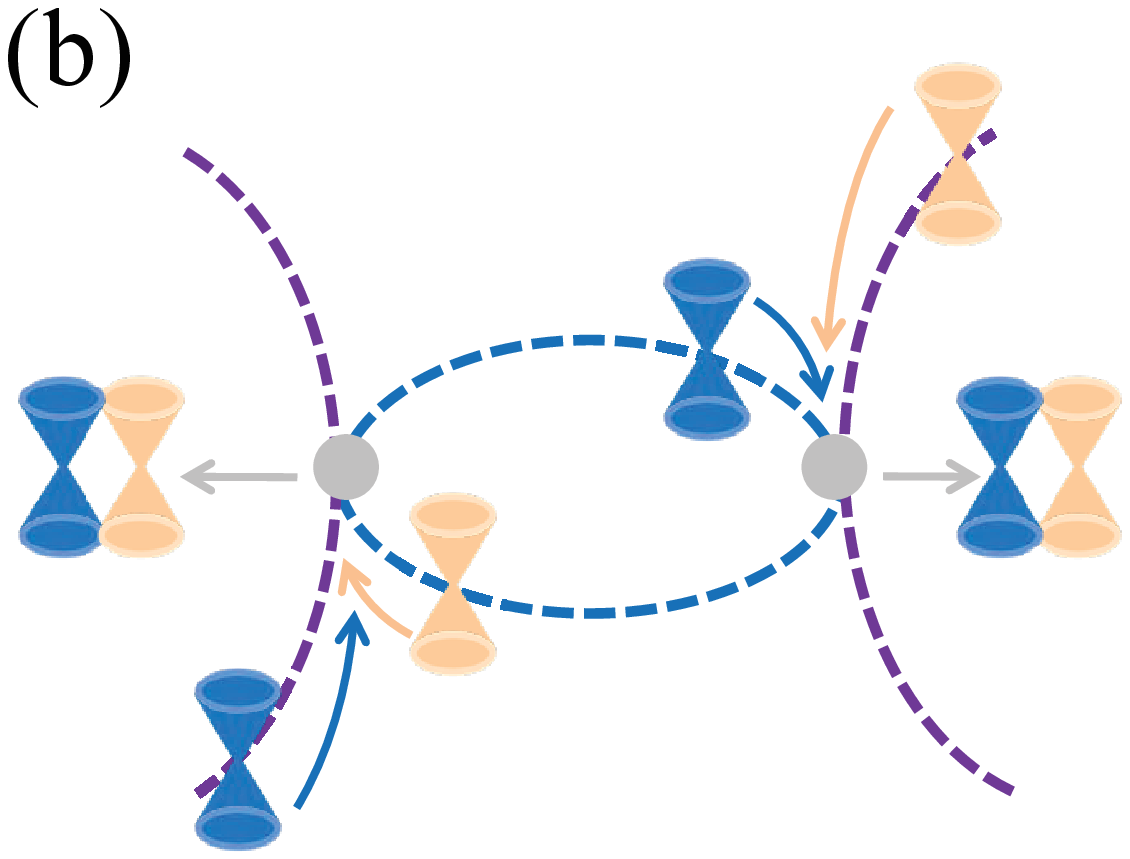}}
\caption{ Illustration of the combination and annihilation process of Weyl points.
The different colors of the Weyl cones stand for different monopole charges.
The incident angle of the light follows a path in great circle passing $(\phi,\theta)=(\pi/2,\pi/4)$
and $(0,\pi/2)$. (a) For the type-I crossing, two Weyl points with
the same monopole charge come close and form a double-Weyl point. (b) For the
type-II crossing, two Weyl points with opposite monopole charge come
close and form a critical gapless point. The arrows indicate how the Weyl points move
with decreasing $\phi$.}  \label{combination}
\end{figure}

Following the standard approach, the electromagnetic coupling is given by $H(\bk)
\rightarrow H(\bk+e\bA(t))$. Since the full Hamiltonian is time-periodic, it can be expanded as
$H(t,\bk)=\sum_{n}H_{n}(\bk)e^{in\omega t}$ with
\begin{eqnarray}
H_{0}(\bk)&=&[\tilde{m}-Bk^{2}]\tau_{x}+ (\lambda k_{y}k_{z}-2D_{1})\tau_{z},\nonumber\\
H_{\pm1}(\bk)&=&-BeA_{0}[(\sin\phi k_{x}-\cos\phi k_{y})\nonumber\\
&&\mp i\eta(\cos\phi\cos\theta k_{x}+\sin\phi\cos\theta k_{y}-\sin\theta k_{z})]\tau_{x}\nonumber\\
&&-\lambda eA_{0}[\cos\phi k_{z}\pm i\eta(\sin\phi\cos\theta k_{z}-\sin\theta k_{y})]\tau_{z}/2,\nonumber\\
H_{\pm2}(\bk)&=&(D_{1}\mp i\eta D_{2})\tau_{z},
\end{eqnarray}
and $H_{n}=0$ for $|n|>2$; $\tilde{m}=m-Be^{2}A_{0}^{2}$,
$D_{1}=\lambda e^{2}A_{0}^{2}\sin\phi\cos\theta\sin\theta/4$, $D_{2}=\lambda e^{2}A_{0}^{2}\cos\phi\sin\theta/4$.
We focus on the off-resonance regimes, and the system is well described by an effective time-independent Hamiltonian,
 which reads\cite{Kitagawa2011,Goldman2014}
\begin{eqnarray}
H_{\rm eff}(\bk)&=&H_{0}+\sum_{n\geq1}\frac{[H_{+n},H_{-n}]}{n\omega}+\mathcal{O}(\frac{1}{\omega^{2}})
\nonumber\\
&=&(\tilde{m}-Bk^{2})\tau_{x}+(\lambda k_{y}k_{z}-2D_{1})\tau_{z}\nonumber\\
&&+\gamma\eta[(\cos\theta k_{z}-\sin\phi\sin\theta k_{y})k_{x}\nonumber\\
&&+\cos\phi\sin\theta(k_{y}^{2}-k_{z}^{2})]\tau_{y}+\cdots,\label{effective}
\end{eqnarray}
where  $\gamma=-2B\lambda e^{2}A_{0}^{2}/\omega$.
Consequently, the energy spectra
of $H_{\rm eff}$ are
\begin{eqnarray}
E_{\pm}(\bk)&=&\pm\left\{(\tilde{m}-Bk^{2})^{2}+(\lambda k_{y}k_{z}-2D_{1})^{2}+\gamma^{2}[(\cos\theta k_{z}\right.\nonumber\\
&&\left.-\sin\phi\sin\theta k_{y})k_{x}+\cos\phi\sin\theta(k_{y}^{2}-k_{z}^{2})]^{2}\right\}^{1/2}.\label{ES}
\end{eqnarray}
For a general incident direction other than $\phi=0$, $\pi$,
and $\theta=0$, $\pi/2$, $\pi$, it is readily found from
Eq.(\ref{ES}) that
there are four Floquet Weyl points. Since $eA_{0}<<\sqrt{m/B}$ under experimental conditions,  the $D_{1}$ term can be neglected because
it only induces a small and trivial change to the positions of the Weyl points. Discarding the $D_{1}$ term, it is straightforward
to determine the positions of the Weyl points, which are at
\begin{eqnarray}
\bQ_{1}&=&-\bQ_{2}=\frac{\sqrt{\tilde{m}/B}}{\sqrt{(\cos\phi\sin\theta)^{2}+\cos^{2}\theta}}(\cos\phi\sin\theta,
0,\cos\theta),\nonumber\\
\bQ_{3}&=&-\bQ_{4}=\sqrt{\tilde{m}/B}(\cos\phi,
\sin\phi,0).\label{position}
\end{eqnarray}
We can expand $H_{\rm eff}$ around these points as
$H_{\alpha=1,2,3,4}(\bq)=\sum_{ij}v_{\alpha,ij}q_{i}\tau_{j}$
with $\bq=\bk-\bQ_{\alpha}$ referring to the momentum relative
to the gapless points. The monopole charge of the Weyl point at $\bQ_{\alpha}$
is simply
$C_{\alpha}= \text{sgn}[\det(v_{\alpha,ij})]$.  A straightforward
calculation gives
\begin{eqnarray}
C_{1}=-C_{2}=C_{3}=-C_{4}=\eta.\label{charge}
\end{eqnarray}
The number of monopole is equal to the number of antimonopole, automatically satisfying
Nielsen-Ninomiya theorem\cite{NIELSEN1981}. From Eq.(\ref{position}) and Eq.(\ref{charge}),
it is readily seen that with the variation of $(\theta,\phi)$ and the handness of
the light, both the positions
and the monopole charges are tunable. Most interestingly,
when the direction $(\phi,\theta)=(0,\pi/2)$ or $(\pi,\pi/2)$ is reached,
we can observe the combination of two Weyl points with the same monopole
charge to form a double-Weyl point. To see this more clearly, notice that when
the light comes in the $x$ direction, i.e., $\phi=0$, $\theta=\pi/2$,
$H_{\rm eff}$ in Eq.(\ref{effective}) reduces to the form of
\begin{eqnarray}
H_{\rm eff}(\bk)=(\tilde{m}-Bk^{2})\tau_{x}+\lambda k_{y}k_{z}\tau_{z}+ \eta\gamma(k_{y}^{2}-k_{z}^{2})\tau_{y},
\end{eqnarray}
which gives two gapless points at $\bQ_{\pm}=\pm(\sqrt{\tilde{m}/B},0,0)$.
A calculation of the Berry-flux number passing through the surface
enclosing $\bQ_+$ or $\bQ_-$ yields the monopole charges, which are \bea C_{\pm}=\pm 2\eta, \eea i.e., they are
double-Weyl points.
A picture illustration of the motion and combination
of Weyl points, as $(\theta,\phi)$ is tuned, is shown in Fig.\ref{combination}(a).

Before closing this section, we briefly discuss crossing nodal lines with cubic symmetry, keeping in mind that
several material candidates of NLSM are found to belong to this class~\cite{Yu2015, Du2016CaTe, Gan2016XB6}.
The cubic symmetry guarantees the existence
of three nodal lines located in mutually orthogonal planes, and the nodal lines
are mutually crossing. Since the crossing is still the type I, the physics of
Floquet double-Weyl points and monopole combination is similar as that of Eq.(\ref{nlsm})
(see Supplemental Material for details).

{\it Monopole annihilation from type-II crossing.---}
Now we turn to the type-II crossing [see Fig.\ref{sketch}(b)], which serves as the key building block of nodal chain\cite{Bzdusek2016,Yu2017chain,wang2017chain}.
The local Hamiltonian near the type-II crossing point can be captured by the following continuum Hamiltonian
\begin{eqnarray}
H(\bk)=[m-B(k_{x}^{2}+k_{y}^{2})+Bk_{z}^{2}]\tau_{x}+\lambda k_{y}k_{z}\tau_{z},
\end{eqnarray}
whose energy spectra are
\begin{eqnarray}
E_{\pm,\bk}=\pm\sqrt{[m-B(k_{x}^{2}+k_{y}^{2})+Bk_{z}^{2}]^{2}+(\lambda k_{y}k_{z})^{2}}.
\end{eqnarray}
Thus, there is a nodal ring at $k_{z}=0$, $k_{x}^{2}+k_{y}^{2}=m/B$, as well as two open nodal lines
of hyperbolic shape at $k_{y}=0$, $k_{x}^{2}-k_{z}^{2}=m/B$. The three nodal lines
touch at two points $\bK_{\pm}=\pm(\sqrt{m/B},0,0)$, which gives the type-II
crossing shown in Fig.\ref{sketch}(b).

Again we consider an incident light described by $\bA(t)=A_{0}(\cos(\omega t)\be_{1}
+\eta\sin(\omega t)\be_{2})$ and follow the procedures of the previous section, we find
that the effective Floquet Hamiltonian takes the form of\cite{monopole}
\begin{eqnarray}
H_{\rm eff}(\bk)&=&[m(\theta)-B(k_{x}^{2}+k_{y}^{2})+Bk_{z}^{2}]\tau_{x}+(\lambda k_{y}k_{z}-2D_{1})\tau_{z}\nonumber\\
&&+\gamma\eta[D_{3}+(\cos\theta k_{z}-\sin\phi\sin\theta k_{y})k_{x}\nonumber\\
&&+\cos\phi\sin\theta(k_{y}^{2}+k_{z}^{2})]\tau_{y},\label{effective2}
\end{eqnarray}
where $m(\theta)=m-Be^{2}A_{0}^{2}\cos^{2}\theta$, and
$D_{3}=-D_{2}Be^{2}A_{0}^{2}\sin^{2}\theta /(\gamma\omega)$.
The $D_{3}$ term is fourth order in
$eA_{0}$, which is small, thus we first neglect it.

For a general incident angle, similar to the type-I case,
the  effect of $D_{1}$ term can be neglected. It is straightforward to
find the four Weyl points at
\begin{eqnarray}
\bQ'_{1}&=&-\bQ'_{2}=
\frac{\sqrt{m(\theta)/B}}{\sqrt{(\cos\phi\sin\theta)^{2}-\cos^{2}\theta}}(-\cos\phi\sin\theta,0,\cos\theta),\nonumber\\
\bQ'_{3}&=&-\bQ'_{4}=
\sqrt{m(\theta)/B}(\cos\phi,\sin\phi,0).\label{position2}
\end{eqnarray}
The corresponding monopole charges are found to be
\begin{eqnarray}
C'_{1}&=&-C'_{2}=\eta,\nonumber\\
C'_{3}&=&-C'_{4}=\eta.\label{charge2}
\end{eqnarray}
Thus, both the positions and the monopole charges of the
Weyl points are highly tunable.
When the incident direction of the light is tuned to the $x$ direction,
i.e., $\phi=0$ and $\theta=\pi/2$,
it is readily seen from Eq.(\ref{position2}) that
$\bQ_{1}'$ and $\bQ_{4}'$ will overlap, similarly for $\bQ_{2}'$ and $\bQ_{3}'$. Eq.(\ref{charge2}) tells us that their monopole charges
are opposite, thus, the annihilation
of two Weyl points with opposite monopole charge will occur, and
a gapless point with $C=0$ is found as the remnant of annihilation.
To be explicit, let us write down $H_{\rm eff}$ for $\phi=0$ and $\theta=\pi/2$:
\begin{eqnarray}
H_{\rm eff}(\bk)=[m-B(k_{x}^{2}+k_{y}^{2})+Bk_{z}^{2}]\tau_{x}+\lambda k_{y}k_{z}\tau_{z}
+\gamma\eta(k_{y}^{2}+k_{z}^{2})\tau_{y}.\quad\quad
\end{eqnarray}
There are only two gapless points, namely $\bK_{\pm}$.  It is readily found
that the monopole charges of $\bK_{\pm}$ are both zero. In fact, the
sign of the coefficient of $\tau_z$ is the same for all $\bk$, preventing a nonzero winding of the pseudospin vector around the origin, thus the monopole charge has to vanish.

Thus, monopole annihilation can be observed using nodal lines with
type-II crossing (Fig.\ref{combination}(b)). Since $\bK_{\pm}$ have
vanishing monopole charge, they are unstable, i.e., they can be gapped out by a perturbation of the
form $\Delta \tau_{z}$ ($\Delta$ denotes a constant).

Now we come back to the effects of the $D_{3}$ term.
With the $D_3$ term, we find that the energy
spectra have a small gap $2|\gamma D_{3}|$ at $\bK_{\pm}$ when the
light comes in the $x$ direction (i.e., a Floquet insulator). Therefore, when
the direction of light is tuned away from the $x$ direction to other directions, the
system undergoes an insulator-WSM transition at certain incident angle, namely, pairs of
Weyl points with opposite monopole charges are created from the Floquet insulators.

{\it Surface state evolution.---}
A key character of Weyl semimetals is the surface Fermi arcs. With the creation of multi-Weyl points, multiple Fermi arcs are naturally expected. We now check it by explicit calculations. Let us focus on the type-I crossing (the similar analysis of type-II crossing is given
in Supplemental Material). We consider that the system occupies the $z>0$ region.
The energy dispersion and the wave functions of the surface states
can be determined by solving the eigenvalue problem $H_{\rm
eff}(k_{x},k_{y},-i\partial_{z})\Psi(x,y,z)=E(k_{x},k_{y})\Psi(x,y,z)$, under
the boundary conditions $\Psi(z=0)=0$ and $\Psi(z\rightarrow+\infty)=0$.  For simplicity, we neglect the $D_{1}$ term at this stage, and take the driving-induced $\tau_{y}$ term
as a perturbation (this is justified as both $D_{1}$ and $\gamma$ are small),
namely, $H_{\rm eff}\simeq H_{0}+\Delta H$ with
\begin{eqnarray}
H_{0}(\bk)&=&(\tilde{m}-Bk^{2})\tau_{x}+\lambda k_{y}k_{z}\tau_{z},\nonumber\\
\Delta H(\bk)&=&\gamma\eta[(\cos\theta k_{z}-\sin\phi\sin\theta k_{y})k_{x}
+\cos\phi\sin\theta(k_{y}^{2}-k_{z}^{2})]\tau_{y}.\quad
\end{eqnarray}
We first solve the eigenfunction $H_{0}(k_{x},k_{y},-i\partial_{z})\Psi(x,y,z)=E_{0}(k_{x},k_{y})\Psi(x,y,z)$, which gives $E_{0}=0$ and
\begin{eqnarray}
\Psi(x,y,z)=\mathcal{N}e^{ik_{x}x}e^{ik_{y}y}(e^{-\kappa_{+}z}-e^{-\kappa_{-}z})\chi,
\end{eqnarray}
with $\mathcal{N}$ a normalization constant, $\chi=(sgn(k_{y}),-i)^{T}/\sqrt{2}$ and
\begin{eqnarray}
\kappa_{\pm}=\frac{\lambda|k_{y}|}{2B}\pm\frac{i}{2B}\sqrt{4B(\tilde{m}-Bk_{x}^{2}-Bk_{y}^{2})-\lambda^{2}k_{y}^{2}}.
\end{eqnarray}
The surface state exists only when $\min\{\text{Re} \kappa_{+},\text{Re} \kappa_{-}\}>0$, i.e.,
$k_{x}^{2}+k_{y}^{2}<\sqrt{\tilde{m}/B}$.

\begin{figure}
\includegraphics[width=7cm, height=5cm]{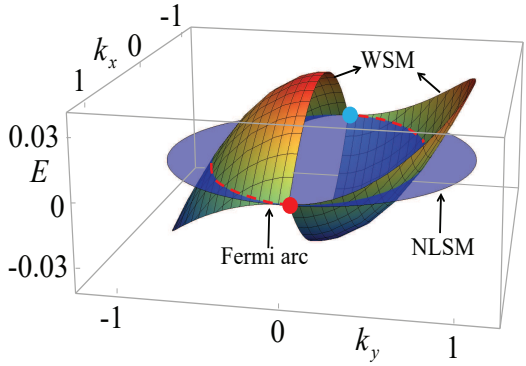}
\caption{ The surface state dispersions of the
pristine crossing-nodal-line semimetal (flat surface) and the Floquet double-Weyl
semimetal (tilted surfaces). The parameters are taken to be $m=B=\lambda=1$,
$\omega=2$, $eA_{0}=0.2$, $\eta=1$, $\phi=0$ and $\theta=\pi/2$.
The filled dots are the
projections of double-Weyl points to the surface Brillouin zone.
The red dashed lines are the two Fermi arcs connecting the two double-Weyl points (for zero chemical potential).  }\label{Fermiarc}
\end{figure}

Now we add the perturbation $\Delta H$, which modifies the dispersion as
\begin{eqnarray}
\Delta E(k_{x},k_{y})&=&\int_{0} ^{\infty}dz\Psi^{\dag}(x,y,z)\Delta H(k_{x},k_{y},-i\partial_{z})\Psi(x,y,z)\nonumber\\
&=&sgn(k_{y})\gamma\eta[(\sin\phi\sin\theta k_{x}k_{y}-\cos\phi\sin\theta k_{y}^{2})\nonumber\\
&&+\kappa_{+}\kappa_{-}\cos\phi\sin\theta].
\end{eqnarray}
Consequently, the surface states of the driven system become dispersive
and the dispersion is given by $E(k_{x},k_{y})=\Delta E(k_{x},k_{y})$ in this perturbation theory.
For the double-Weyl point case, i.e., $\theta=\pi/2$ and $\phi=0$, the energy dispersion reads
\begin{eqnarray}
E(k_{x},k_{y})={\rm sgn}(k_{y})\gamma\eta[\tilde{m}/B-k_{x}^{2}-2k_{y}^{2}].
\end{eqnarray}
The surface state dispersions of
the pristine crossing-nodal-line semimetal and the Floquet double-Weyl
semimetal are shown in Fig.\ref{Fermiarc}. It is readily seen
that the driving tears and tilts the flat drumhead surface band of the pristine NLSM, giving rise to two Fermi arcs. The number of Fermi arcs is equal
to the monopole charge of the Weyl points.

{\it Effect of spin-orbit coupling.---}
Now we discuss the effect of spin-orbit coupling (SOC).
As long as the pristine crossing structure is robust against SOC,
such as that of the proposed material IrF$_{4}$\cite{Bzdusek2016}, where it is protected by nonsymmorphic symmetries,
the introduction of SOC will only induce a change in the positions of the
Flqouet Weyl points.
On the other hand, if the pristine crossing structure is fragile to SOC, such as that of the candidate CaTe, where the nodal lines are predicated to evolve into
Dirac points in the presence of SOC\cite{Du2016CaTe}, we find the Floquet double-Weyl points become
unstable and will be split into Floquet single-Weyl points (Supplemental Material).

{\it Experimental estimations.---}Among other approaches, an optimal
experimental method to verify this proposal is the pump-probe angle-resolved photoemission spectroscopy
(ARPES)\cite{wang2013observation,mahmood2016selective, Giovannini2016floquet},
which can directly measure the locations of double-Weyl points. Another approach is
to measure the incident-angle-dependent Hall voltage, which is determined by the locations of the Floquet
Weyl points\cite{Chan2016hall,Yan2016tunable}. Here we provide an estimation based on the material candidate
Cu$_{3}$NPd\cite{Kim2015,Yu2015}. Under
the experimental condition of ref.\cite{wang2013observation}, $\gamma$ is estimated to be
of the order of $0.1\lambda$,  and a film sample with size
$l_{x}\times l_{y}\times d= 100$$\mu$m$\times100$$\mu$m$\times500$nm  can
generate an incident-angle-dependent Hall voltage of the order of $20$ mV if a
dc current of $100$ mA is applied in $y$ direction (Supplemental material), well within the capacity of current experiments.

{\it Conclusions.---}There have been extensive theoretical and experimental studies of Weyl points
with monopole charge $\pm 1$, however,  multi-Weyl semimetals have not been
well studied so far due to the lack of materials. Here, we show that multi-Weyl
points can be realized in driven nodal-line semimetals with novel line-connectivity
(crossing nodal lines and nodal chains). In addition to suggesting a way to realize
multi-Weyl semimetals, this work indicates that novel nodal lines are versatile platforms
in the field of topological semimetals. Our proposal may also
be generalized to cold-atom systems where periodic driving can be realized
by shaking the optical lattice\cite{jotzu2014experimental,parker2013direct,Zheng2014}.

Note added:  Upon finishing this manuscript, we become aware of a related preprint\cite{Ezawa2017Photoinduced}, in which type-I crossing is studied.

{\it Acknowledgements.---}
We would like to thank Gang Chen and Ling Lu for useful discussions.
This work is supported by NSFC (No. 11674189). Z. Y. is supported in part by China Postdoctoral Science
Foundation (No. 2016M590082).

\bibliography{dirac}

\vspace{8mm}

{\bf Supplemental Material}

\vspace{4mm}

This supplemental material contains: (i) The derivation of
the effective Floquet Hamiltonian for the type-II crossing model.
(ii) Floquet Weyl points in crossing-nodal-line semimetals with cubic symmetry.
(iii) Surface state evolution. (vi) Effect of spin-orbit coupling. (v) Experimental estimation.

\section{Derivation of the effective Floquet Hamiltonian
for the type-II crossing model}
\label{derivation}

The starting Hamiltonian is
\begin{eqnarray}
H(\bk)=[m-B(k_{x}^{2}+k_{y}^{2})+Bk_{z}^{2}]\tau_{x}+\lambda k_{y}k_{z}\tau_{z}.
\end{eqnarray}
We consider an incident light in the direction $\bn=(\cos\phi\sin\theta,\sin\phi\sin\theta,\cos\theta)$.
The  vector potential of the light is $\bA(t)=A_{0}(\cos(\omega t)\be_{1}
+\eta\sin(\omega t)\be_{2})$, with $\eta=\pm1$ corresponding to the right-handed
and left-handed circularly polarized light, respectively. Here, $\be_{1}=(\sin\phi,-\cos\phi,0)$
and $\be_{2}=(\cos\phi\cos\theta,\sin\phi\cos\theta,-\sin\theta)$ are two
vectors perpendicular to $\bn$, satisfying $\be_{1}\cdot\be_{2}=0$.

The electromagnetic coupling is given by $H(\bk)
\rightarrow H(\bk+e\bA(t))$. The full Hamiltonian is time-periodic, therefore, it can be expanded as
$H(t,\bk)=\sum_{n}H_{n}(\bk)e^{in\omega t}$ with
\begin{eqnarray}
H_{0}(\bk)&=&[m(\theta)-B(k_{x}^{2}+k_{y}^{2})+Bk_{z}^{2}]\tau_{x}
+ (\lambda k_{y}k_{z}-2D_{1})\tau_{z},\nonumber\\
H_{\pm1}(\bk)&=&-BeA_{0}[(\sin\phi k_{x}-\cos\phi k_{y})\nonumber\\
&&\mp i\eta(\cos\phi\cos\theta k_{x}+\sin\phi\cos\theta k_{y}+\sin\theta k_{z})]\tau_{x}\nonumber\\
&&-\lambda eA_{0}[\cos\phi k_{z}\pm i\eta(\sin\phi\cos\theta k_{z}-\sin\theta k_{y})]\tau_{z}/2,\nonumber\\
H_{\pm2}(\bk)&=&D_{0}\tau_{x}+(D_{1}\mp i\eta D_{2})\tau_{z},
\end{eqnarray}
where $m(\theta)=m-Be^{2}A_{0}^{2}\cos^{2}\theta$, and $D_{0}=-Be^{2}A_{0}^{2}\sin^{2}\theta/2$,
$D_{1}=\lambda e^{2}A_{0}^{2}\sin\phi\cos\theta\sin\theta/4$, and $D_{2}=\lambda e^{2}A_{0}^{2}\cos\phi\sin\theta/4$.

When $\omega$ is in the off-resonance regime, the system is well described by an effective time-independent Hamiltonian,
which reads
\begin{eqnarray}
H_{\rm eff}(\bk)&=&H_{0}+\sum_{n\geq1}\frac{[H_{+n},H_{-n}]}{n\omega}+\mathcal{O}(\frac{1}{\omega^{2}})
\nonumber\\
&=&[m(\theta)-B(k_{x}^{2}+k_{y}^{2})+Bk_{z}^{2}]\tau_{x}+(\lambda k_{y}k_{z}-2D_{1})\tau_{z}\nonumber\\
&&+\gamma\eta[D_{3}+(-\sin\phi\sin\theta k_{y}+\cos\theta k_{z})k_{x}\nonumber\\
&&+\cos\phi\sin\theta(k_{y}^{2}+k_{z}^{2})]\tau_{y}+\cdots,\label{floquet}
\end{eqnarray}
where $m(\theta)=m-Be^{2}A_{0}^{2}\cos^{2}\theta$, $\gamma=-2B\lambda(eA_{0})^{2}/\omega$, and
$D_{3}=-D_{2}Be^{2}A_{0}^{2}\sin^{2}\theta /(\gamma\omega)$.

\section{Crossing nodal lines with cubic symmetry}\label{CSHamiltonian}

\begin{figure}
\subfigure{\includegraphics[width=6cm, height=5.5cm]{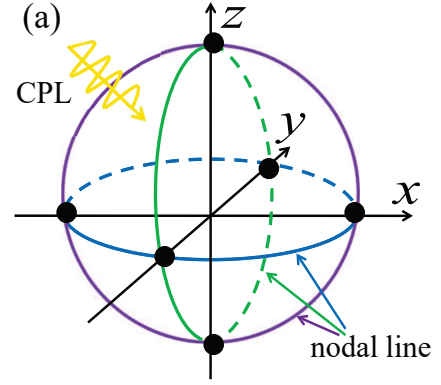}}
\caption{ Three nodal lines are located in three mutually orthogonal planes,
and any two of them are mutually intersected.}  \label{cubic}
\end{figure}

The Hamiltonian for NLSMs with cubic symmetry
is given by\cite{Kim2015, Yu2015}
\begin{eqnarray}
H(\bk)=(m-Bk^{2})\tau_{x}+\lambda k_{x}k_{y}k_{z}\tau_{z}.
\end{eqnarray}
The energy spectra read
\begin{eqnarray}
E_{\pm}(\bk)=\pm\sqrt{(m-Bk^{2})^{2}+(\lambda k_{x}k_{y}k_{z})^{2}}
\end{eqnarray}
There are three nodal lines located in the three mutually perpendicular planes, respectively, i.e.,
the $k_{x}=0$ plane, the $k_{y}=0$ plane, and the $k_{z}=0$ plane.
Any two of the nodal lines mutually cross, as shown in Fig.\ref{cubic}

Let us consider that the system is driven by a circularly polarized light with $\bA(t)=A_{0}(\cos(\omega t)\be_{1}
+\eta\sin(\omega t)\be_{2})$. The full time-periodic Hamiltonian
can be expanded as $H(t,\bk)=\sum_{n}H_{n}(\bk)e^{in\omega t}$ with
\begin{eqnarray}
H_{0}(\bk)&=&[m-Be^{2}A_{0}^{2}-Bk^{2}]\tau_{x}+ (\lambda k_{x}k_{y}k_{z}-G_{1})\tau_{z},\nonumber\\
H_{\pm1}(\vec{k})&=&-BeA_{0}[(\sin\phi k_{x}-\cos\phi k_{y})\nonumber\\
&&\mp i\eta(\cos\phi\cos\theta k_{x}+\sin\phi\cos\theta k_{y}-\sin\theta k_{z})]\tau_{x}\nonumber\\
&&+\lambda eA_{0}\{[(\sin\phi k_{y}k_{z}-\cos\phi k_{x}k_{z})\mp i\eta(\cos\phi\cos\theta k_{y}k_{z}\nonumber\\
&&+\sin\phi\cos\theta k_{x}k_{z}-\sin\theta k_{x}k_{y})]+G_{2}\pm iG_{3}\}\tau_{z}/2,\nonumber\\
\end{eqnarray}
where
\begin{eqnarray}
G_{1}&=&\frac{\lambda (eA_{0})^{2}}{4}(k_{x}\sin\phi\sin2\theta+k_{y}\cos\phi\sin2\theta+k_{z}\sin2\phi\sin^{2}\theta),\nonumber\\
G_{2}&=&\frac{\lambda(eA_{0})^{3}}{8}[2\cos^{2}\phi-\eta\cos2\phi]\sin2\theta,\nonumber\\
G_{3}&=&\frac{\eta\lambda(eA_{0})^{3}}{8}[2\cos^{2}\theta\sin\theta-\sin^{3}\theta]\sin2\phi.
\end{eqnarray}
All $H_{\pm n}$ with $n>1$ will not be  given explicitly  because they
contain only one pauli matrix $\tau_{z}$, thus, they do not contribute
to the effective Hamiltonian, which involves commutators. Following the approach in Eq.(\ref{floquet}),
we obtain the effective Hamiltonian:
\begin{eqnarray}
H_{\rm eff}(\bk)
&=&[\tilde{m}-Bk^{2}]\tau_{x}+ (\lambda k_{x}k_{y}k_{z}-G_{1})\tau_{z}\nonumber\\
&&+\gamma\eta[\cos\theta k_{z}(k_{x}^{2}-k_{y}^{2})+\sin\phi\sin\theta k_{y}(k_{z}^{2}-k_{x}^{2})\nonumber\\
&&+\cos\phi\sin\theta k_{x}(k_{y}^{2}-k_{z}^{2})]\tau_{y}+G_{4}\tau_{y}\cdots,
\end{eqnarray}
where $\tilde{m}=m-Be^{2}A_{0}^{2}$, $\gamma=-2B\lambda(eA_{0})^{2}/\omega$, and
\begin{eqnarray}
G_{4}&=&\frac{2BeA_{0}}{\omega}[G_{3}(\sin\phi k_{x}-\cos\phi k_{y})\nonumber\\
&&-\eta G_{2}(\cos\phi\cos\theta k_{x}+\sin\phi\cos\theta k_{y}-\sin\theta k_{z})].
\end{eqnarray}
To simplify the discussion, we take into account the fact that $eA_{0}<<\sqrt{m/B}$, so that for a general incident
direction, both $G_{1}$ and $G_{4}$ can be safely neglected.
Under this approximation, the energy spectra read
\begin{eqnarray}
E_{\pm}(\bk)&=&\pm\{(\tilde{m}-Bk^{2})^{2}+(\lambda k_{x}k_{y}k_{z})^{2}+\gamma^{2}[\cos\theta k_{z}(k_{x}^{2}-k_{y}^{2})\nonumber\\
&&+\sin\phi\sin\theta k_{y}(k_{z}^{2}-k_{x}^{2})+\cos\phi\sin\theta k_{x}(k_{y}^{2}-k_{z}^{2})]^{2}\}^{1/2}.\nonumber\\
\end{eqnarray}
It is readily found that there are six Floquet Weyl points
when $\theta\neq0$, $\pi$, or
$\{\phi,\theta\}\neq\{\{0,\pi/2,\pi,3\pi/2\},\{\pi/2\}\}$,
\begin{eqnarray}
\bQ_{1}&=&-\bQ_{2}=\frac{\sqrt{\tilde{m}/B}}{\sqrt{(\sin\phi\sin\theta)^{2}+\cos^{2}\theta}}(0,
\sin\phi\sin\theta,\cos\theta),\nonumber\\
\bQ_{3}&=&-\bQ_{4}=\frac{\sqrt{\tilde{m}/B}}{\sqrt{(\cos\phi\sin\theta)^{2}+\cos^{2}\theta}}(\cos\phi\sin\theta,
0,\cos\theta),\nonumber\\
\bQ_{5}&=&-\bQ_{6}=\sqrt{\tilde{m}/B}(\cos\phi,
\sin\phi,0),\label{positions}
\end{eqnarray}
and their monopole charge are given by
\begin{eqnarray}
C_{1}=-C_{2}=C_{3}=-C_{4}=C_{5}=-C_{6}=\eta.\label{charges}
\end{eqnarray}
When $\theta=0$, $\pi$, or
$\{\phi,\theta\}=\{\{0,\pi/2,\pi,3\pi/2\},\{\pi/2\}\}$, one of the three
nodal lines remains, and the other two become a pair of double-Weyl points.
For example, it is readily seen from Eq.(\ref{positions}) that when $(\phi,\theta)$ is
tuned to $(0,\pi/2)$, $Q_{3}$ and $Q_{5}$ ($Q_{4}$ and $Q_{6}$) will come close to each other; when $(\phi,\theta)$ is
tuned to $(\pi/2,\pi/2)$, $Q_{1}$ and $Q_{5}$ ($Q_{2}$ and $Q_{6}$) will come close to each other;
when $\theta$ is tuned to $0$ or $\pi$, $Q_{1}$ and $Q_{3}$ ($Q_{2}$ and $Q_{4}$) will come close to each other.

Without loss of generality, we consider the case $(\phi,\theta)=(0,\pi/2)$ to see the monopole combination. For this special case, both $G_{1}$ and $G_{4}$ are strictly
equal to zero, and $H_{\rm eff}$ reduces to
\begin{eqnarray}
H_{\rm eff}=[\tilde{m}-Bk^{2}]\tau_{x}+ \lambda k_{x}k_{y}k_{z}\tau_{z}
+\gamma\eta k_{x}(k_{y}^{2}-k_{z}^{2})\tau_{y}.
\end{eqnarray}
which gives two double-Weyl points at $Q_{\pm}=\pm(\sqrt{\tilde{m}/B},0,0)$
with monopole charge \bea C_{\pm}=\pm 2\eta. \eea  Besides the two double-Weyl points,
$H_{\rm eff}$ also gives a nodal line which is located in the $k_{x}=0$ plane and
determined by $k_{y}^{2}+k_{z}^{2}=\tilde{m}/B$. The survival of this nodal line
originates from the fact that the incident direction of the light is perpendicular to the plane in
which the nodal line is located.
The appearance of this additional nodal line
does not affect the combination of Weyl points with the same monopole
charge to form double-Weyl points.

\section{Surface state evolution of Type-II crossing}

The effective Hamiltonian of the type-II crossing is given by (see Eq.(\ref{floquet}))
\begin{eqnarray}
H_{\rm eff}(\bk)
&=&[m(\theta)-B(k_{x}^{2}+k_{y}^{2})+Bk_{z}^{2}]\tau_{x}+(\lambda k_{y}k_{z}-2D_{1})\tau_{z}\nonumber\\
&&+\gamma\eta[D_{3}+(-\sin\phi\sin\theta k_{y}+\cos\theta k_{z})k_{x}\nonumber\\
&&+\cos\phi\sin\theta(k_{y}^{2}+k_{z}^{2})]\tau_{y}.
\end{eqnarray}
We consider that the system occupies the whole $z>0$ region. Similar to the procedures
in the main article, we neglect the $D_{1}$ term and $D_{3}$ term, and take the driving-induced $\tau_{y}$ term
as a perturbation,
i.e., $H_{\rm eff}\simeq H_{0}+\Delta H$ with
\begin{eqnarray}
H_{0}(\bk)&=&[m(\theta)-B(k_{x}^{2}+k_{y}^{2})+Bk_{z}^{2}]\tau_{x}+\lambda k_{y}k_{z}\tau_{z}\nonumber\\
\Delta H(\bk)&=&\gamma\eta[(-\sin\phi\sin\theta k_{y}+\cos\theta k_{z})k_{x}\nonumber\\
&&+\cos\phi\sin\theta(k_{y}^{2}+k_{z}^{2})]\tau_{y}.
\end{eqnarray}
Solving the eigenfunction $H_{0}(k_{x},k_{y},-i\partial_{z})\Psi(x,y,z)=E_{0}(k_{x},k_{y})\Psi(x,y,z)$
under the boundary conditions $\Psi(z=0)=0$ and $\Psi(z\rightarrow+\infty)=0$
gives $E_{0}=0$ and
\begin{eqnarray}
\Psi(x,y,z)=\mathcal{N}e^{ik_{x}x}e^{ik_{y}y}(e^{-\kappa_{+}z}-e^{-\kappa_{-}z})\chi
\end{eqnarray}
with $\chi=(sgn(k_{y}),i)^{T}/\sqrt{2}$ and
\begin{eqnarray}
\kappa_{\pm}=\frac{\lambda|k_{y}|}{2B}\pm\frac{i}{2B}\sqrt{4B(\tilde{m}(\theta)-Bk_{x}^{2}-Bk_{y}^{2})-\lambda^{2}k_{y}^{2}},
\end{eqnarray}
The surface state exists only when $\min\{\text{Re} \kappa_{+},\text{Re} \kappa_{-}\}>0$. Here,
$\mathcal{N}$ is a normalization constant, which takes the form of
\begin{eqnarray}
\mathcal{N}=\left\{\begin{array}{cc}
              \sqrt{-\frac{2\kappa_{+}\kappa_{-}(\kappa_{+}+\kappa_{-})}{(\kappa_{+}-\kappa_{-})^{2}}},  & \text{for}\, \kappa_{+}=\kappa_{-}^{*}, \\
              \sqrt{\frac{2\kappa_{+}\kappa_{-}(\kappa_{+}+\kappa_{-})}{(\kappa_{+}-\kappa_{-})^{2}}}, & \text{for}\, \kappa_{\pm}=\kappa_{\pm}^{*}\, \text{and}\, \kappa_{\pm}>0,
            \end{array}\right.
\end{eqnarray}
The modification to the energy dispersion of the surface states
by $\Delta H$ is
\begin{eqnarray}
\Delta E(k_{x},k_{y})&=&\int_{0} ^{\infty}dz\Psi^{\dag}(x,y,z)\Delta H(k_{x},k_{y},-i\partial_{z})\Psi(x,y,z)\nonumber\\
&=&-sgn(k_{y})\gamma\eta[(\sin\phi\sin\theta k_{x}k_{y}-\cos\phi\sin\theta k_{y}^{2})\nonumber\\
&&-\kappa_{+}\kappa_{-}\cos\phi\sin\theta].
\end{eqnarray}
For $\theta=\pi/2$ and $\phi=0$, namely, the angle corresponding to monopole annihilation,
the energy dispersion is
\begin{eqnarray}
E(k_{x},k_{y})={\rm sgn}(k_{y})\gamma\eta[\tilde{m}/B-k_{x}^{2}].
\end{eqnarray}
For this angle, the surface state only exists in the regime $\tilde{m}/B>k_{x}^{2}+k_{y}^{2}$.
It is immediately seen that Fermi arc is absent at the Fermi energy $E_{F}=0$,
agreeing with the fact that monopoles have annihilated with
each other at this angle.

\section{Effect of spin-orbit coupling}
\subsection{Crossing nodal lines robust against spin-orbit coupling}

When materials has certain symmetry, e.g., mirror symmetry or nonsymmorphic symmetry,
nodal lines can stably exist even
in the presence of spin-orbital coupling\cite{Bian2015nodal, Bian2015TlTaSe, Bzdusek2016}. For instance, we assume that the
crossing nodal lines are around a high symmetric point (most of the predicted materials
fall into this class) and the low-energy effective Hamiltonian is given by
\begin{eqnarray}
H(\bk)=(m-Bk^{2})\tau_{x}+\lambda k_{x}k_{y}k_{z}\tau_{z}+\lambda_{so}\tau_{x}\sigma_{z},
\end{eqnarray}
where $\lambda_{so}$ denotes the spin-orbit coupling strength.  For this type of spin-orbit coupling, it only
induces a change of the size of the nodal lines, but does not destroy the crossing
structure.

We consider that a CPL is incident in $x$ direction and described by
the vector potential $\bA=A_{0}(0,\cos\omega t, \eta\sin\omega t)$.
Following the same steps as in the main article,  we obtain the effective
Hamiltonian in the off-resonant regime, which is
\begin{eqnarray}
H_{\rm eff}(\bk)&=&(m-Bk^{2})\tau_{x}+\lambda k_{x}k_{y}k_{z}\tau_{z}+\lambda_{so}\tau_{x}\sigma_{z}\nonumber\\
&&+\gamma\eta k_{x}(k_{y}^{2}-k_{z}^{2})\tau_{y}.
\end{eqnarray}
It is readily found that there are two pairs of double-Weyl points, with one pair
located at $W_{1,\pm}=\pm(\sqrt{(\tilde{m}+\lambda_{so})/B},0,0)$, and the other
pair located at $W_{2,\pm}=\pm(\sqrt{(\tilde{m}-\lambda_{so})/B},0,0)$. Thus,
when the crossing-nodal-line structure are robust against the spin-orbit coupling,
the double-Weyl points can still be dynamically created. The effect of spin-orbit coupling
is to induce a shift of the positions of the Floquet Weyl points.

\subsection{Crossing nodal lines not robust against spin-orbit coupling}

The nodal lines in some of the predicted material candidates evolve into
pairs of Dirac points in the presence of spin-orbit coupling. To describe this case,
we consider a simplified model related to the crossing-nodal-line semimetal CaTe\cite{Du2016CaTe}.
\begin{eqnarray}
H(\bk)=(m-Bk^{2})\tau_{x}+\lambda k_{x}k_{y}k_{z}\tau_{z}\sigma_{z}
+\lambda_{so}\tau_{z}(k_{x}\sigma_{y}-k_{y}\sigma_{x}).\quad
\end{eqnarray}
Without the spin-orbit coupling term, i.e., $\lambda_{so}=0$, the Hamiltonian hosts three mutually orthogonal
nodal lines. The presence of the spin-orbit coupling term will gap out the nodal lines, leaving only
two Dirac points at $(0, 0, \pm\sqrt{m/B})$.

\begin{figure}[t]
\subfigure{\includegraphics[width=7cm, height=6cm]{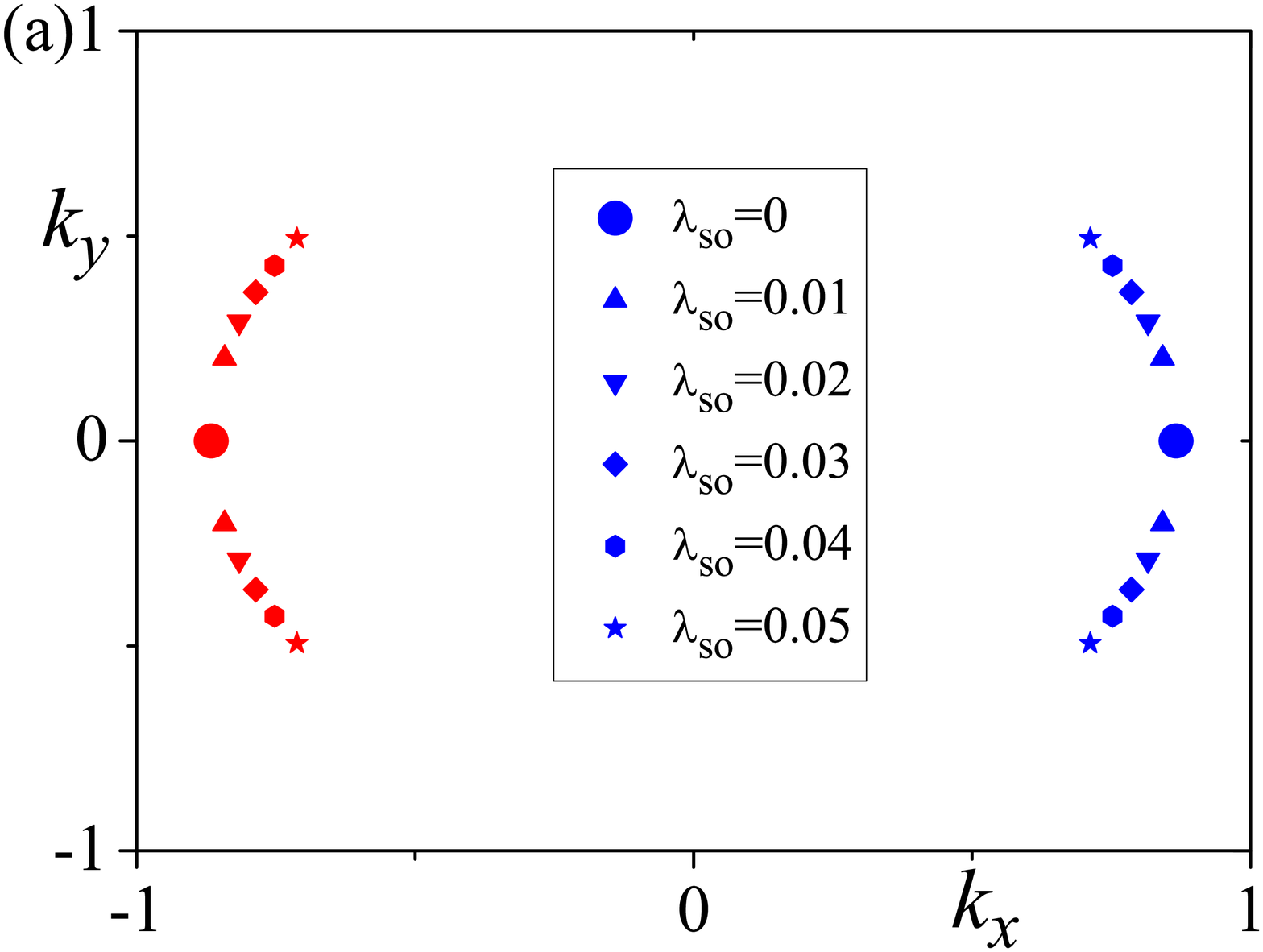}}
\subfigure{\includegraphics[width=7cm, height=6cm]{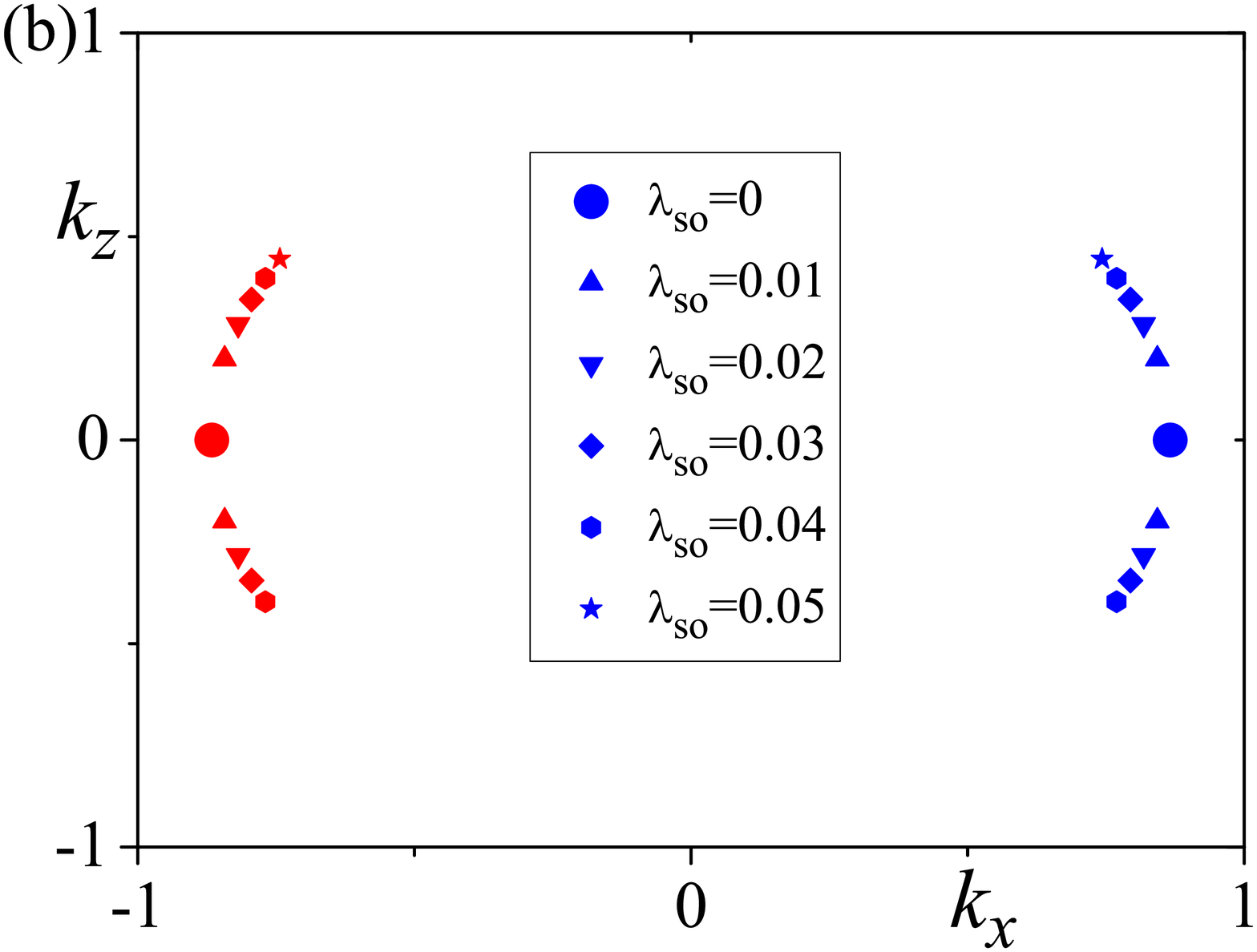}}
\caption{The splitting of doubly-degenerate double-Weyl points into Weyl points in the presence
of spin-orbit coupling. The parameters
chosen for illustration are: $m=1$, $B=1$, $\lambda=1$, $eA_{0}=0.5$, $\omega=2$.
(a) $k_{z}=0$ plane. (b) $k_{y}=0$ plane. The two filled dots mark the locations of the double-Weyl points in the absence of spin-orbit coupling,
and other symbols denote the location of single-Weyl points in the presence of spin-orbit coupling. Red and blue color represent opposite monopole charges.   }\label{split}
\end{figure}

Now we also consider a CPL is incident in $x$ direction and described by the vector
potential $\bA=A_{0}(0,\cos\omega t, \eta\sin\omega t)$.
The effective Hamiltonian can be similarly obtained, which is
\begin{eqnarray}
H_{\rm eff}(\bk)&=&(\tilde{m}-Bk^{2})\tau_{x}+\lambda k_{x}k_{y}k_{z}\tau_{z}\sigma_{z}+\lambda_{so}\tau_{z}(k_{x}\sigma_{y}-k_{y}\sigma_{x})\nonumber\\
&&+\gamma'\eta[2B\lambda k_{x}(k_{y}^{2}-k_{z}^{2})\tau_{y}\sigma_{z}+2B\lambda_{so}k_{z}\tau_{y}\sigma_{x}\nonumber\\
&&+\lambda\lambda_{so}k_{x}k_{y}\sigma_{y}],
\end{eqnarray}
where $\gamma'=-(eA_{0})^{2}/\omega$. When $\lambda_{so}=0$, $H_{\rm eff}(\bk)$ reduces
to
\begin{eqnarray}
H_{\rm eff}(\bk)=(\tilde{m}-Bk^{2})\tau_{x}+\lambda k_{x}k_{y}k_{z}\tau_{z}\sigma_{z}+2B\lambda \gamma'\eta k_{x}(k_{y}^{2}-k_{z}^{2})\tau_{y}\sigma_{z},\nonumber
\end{eqnarray}
which harbors a pair of doubly-degenerate double-Weyl points at $Q_{\pm}=\pm(\sqrt{\tilde{m}/B},0,0)$.
When $\lambda_{so}\neq0$,
the energy spectra of this Hamiltonian
can not be analytically solved, thus we calculate it numerically. As shown in
Fig.\ref{split}, the pair of doubly-degenerate double-Weyl points are spilt into four pairs of
Weyl points in the presence of weak spin-orbit coupling, with two pairs located at
the $k_{y}=0$ plane, and the other two pairs located at the $k_{z}=0$ plane.

\section{Experimental estimations}

\subsection{Estimation of the modification to energy bands}

The modification to energy bands by the driving can be evaluated by calculating
the quantity $|\gamma/\lambda|=2Be^{2}A_{0}^{2}/\omega$, in which $\lambda$ is the parameter of static Hamiltonian (see main article). As $A_{0}=\mathcal{E}_{0}/\omega$,
where $\mathcal{E}_{0}$ is the electric field strength, the quantity can be further
rewritten as
\begin{eqnarray}
|\frac{\gamma}{\lambda}|=\frac{2Be^{2}\mathcal{E}_{0}^{2}}{\omega^{3}}.
\end{eqnarray}
For $\omega$ and $\mathcal{E}_{0}$, we adopt the experimental parameters in
the pump-probe experiment\cite{wang2013observation}, where $\omega=120$ meV,
and  $\mathcal{E}_{0}=2.5\times10^{7}$ V/m. For $B$, we take the material candidate
Cu$_{3}$NPd to make an estimate. According to the band structure obtained by first
principle calculation\cite{Kim2015,Yu2015}, $m\sim0.5$ eV, and $\sqrt{m/B}\sim0.2\pi/a$ with $a=3.85$ {\AA}\cite{Weber1996Cu3NPd},
thus $B\sim2\times10^{-19}$ eVm$^2$. Then
\begin{eqnarray}
|\frac{\gamma}{\lambda}|&\sim&\frac{2\times2\times10^{-19}eVm^{2}\times (2.5\times10^{7}V/m)^{2}}{(0.12eV)^{3}}\nonumber\\
&\approx&0.14.
\end{eqnarray}
Such a  magnitude of modification can be readily observed in current experiments.

\subsection{Estimation of Hall voltage in experiments}

Due to the existence of monopole charges, anomalous Hall
effect will show up in WSMs. At zero temperature and neutrality point (chemical potential $\mu=0$), the Hall conductivities are
given by\cite{yang2011}
\begin{eqnarray}
\sigma_{\alpha\beta}=\frac{e^{2}}{h}\epsilon^{\alpha\beta\tau}\sum_{i}\frac{C_{i}k_{\tau}^{(i)}}{2\pi},\label{formula}
\end{eqnarray}
where $\alpha, \beta, \tau=\{x,y,z\}$ and  $\epsilon^{\alpha\beta\tau}$ is the Levi-Civita symbol;
$C_{i}$ and $k_{\tau}^{(i)}$ denotes the monopole charge and the $\tau$-component of the momentum of the $i$-th Weyl point,
respectively.

Now we consider the NLSM Hamiltonian with cubic symmetry (see Sec.\ref{CSHamiltonian}).
For a CPL incident in a general direction $\bn=(\cos\phi\sin\theta,\sin\phi\sin\theta,\cos\theta)$,
the Hall conductivities can be obtained according to Eq.(\ref{formula}), which are
\begin{eqnarray}
\sigma_{xy}&=&\eta\frac{e^{2}}{h}\frac{\sqrt{\tilde{m}/B}}{\pi}\{\frac{\cos\theta[1-\delta_{\theta,\pi/2}(\delta_{\phi,0}+\delta_{\phi,\pi})]}
{\sqrt{(\sin\phi\sin\theta)^{2}+\cos^{2}\theta}}\nonumber\\
&&+\frac{\cos\theta[1-\delta_{\theta,\pi/2}(\delta_{\phi,\pi/2}+\delta_{\phi,3\pi/2})]}{\sqrt{(\cos\phi\sin\theta)^{2}+\cos^{2}\theta}}\},\nonumber\\
\sigma_{yz}&=&\eta\frac{e^{2}}{h}\frac{\sqrt{\tilde{m}/B}}{\pi}
\{\frac{\cos\phi\sin\theta[1-\delta_{\theta,\pi/2}(\delta_{\phi,\pi/2}+\delta_{\phi,3\pi/2})]}{\sqrt{(\cos\phi\sin\theta)^{2}+\cos^{2}\theta}}\nonumber\\
&&+\cos\phi(1-\delta_{\theta,0}-\delta_{\theta,\pi})\},\nonumber\\
\sigma_{zx}&=&\eta\frac{e^{2}}{h}\frac{\sqrt{\tilde{m}/B}}{\pi}
\{\frac{\sin\phi\sin\theta[1-\delta_{\theta,\pi/2}(\delta_{\phi,0}+\delta_{\phi,\pi})]}
{\sqrt{(\sin\phi\sin\theta)^{2}+\cos^{2}\theta}}\nonumber\\
&&+\sin\phi(1-\delta_{\theta,0}-\delta_{\theta,\pi})\}.\label{AHE}
\end{eqnarray}

\begin{figure}[t]
\includegraphics[width=7cm, height=4cm]{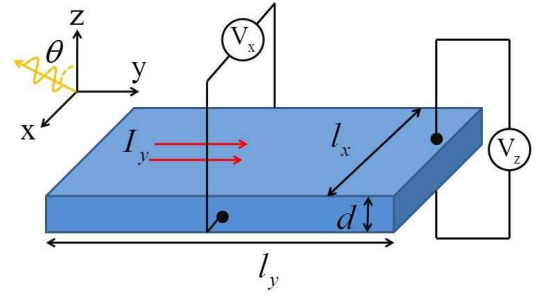}
\caption{Schematic picture of the experimental setup for the measurement of angle-dependent
anomalous Hall effect. The CPL is incident in the $x$-$z$ plane. $I_{y}$ represents an electric current in the $y$ direction, $l_{y}$
denotes the spacing between current contacts. $l_{x}$ and $d$ denote the length in the $x$ direction and the thickness in the $z$ direction (also the spacing between voltage contacts). $V_{x}$ and
$V_{z}$ are the Hall voltages to be measured. }\label{setup}
\end{figure}
Now we consider an experimental setup illustrated in Fig.\ref{setup}.
In Fig.\ref{setup}, $I_{y}$ represents the current in $y$ direction and
we will assume that it is  distributed uniformly. $l_{y}$ denotes the spacing
between current contacts, while $l_{x}$ and $d$ denote the length in the $x$ direction and the thickness in the $z$ direction (also the spacing between voltage contacts). Because
the current is in the $y$ direction, $\sigma_{zx}$ will play no role in transport,
so we restrict the CPL in the $x$-$z$ plane in which $\sigma_{zx}$ naturally vanishes.

The resultant Hall voltage can be estimated as follows:
\begin{eqnarray}
V_{x}&=&R_{xy}I_{y}\nonumber\\
&\approx&-\frac{\sigma_{xy}\delta/d}{\sigma_{yy}^{2}+(\sigma_{xy}\delta/d)^{2}+(\sigma_{yz}\delta/l_{x})^{2}}
\times\frac{l_{x}}{l_{y}d}\times I_{y},\nonumber\\
V_{z}&=&R_{zy}I_{y}\nonumber\\
&\approx&\frac{\sigma_{yz}\delta/l_{x}}{\sigma_{yy}^{2}+(\sigma_{xy}\delta/d)^{2}+(\sigma_{yz}\delta/l_{x})^{2}}
\times\frac{d}{l_{x}l_{y}}\times I_{y}.\label{voltage}
\end{eqnarray}
Here we have assumed that $\sigma_{xx}=\sigma_{yy}=\sigma_{zz}$ for simplicity
as the original static system has cubic symmetry. $\delta$ is the penetration
depth, determined by $\delta(\omega)=\frac{n(\omega)\epsilon_{0}c}{\text{Re}\sigma(\omega)}$,
with $n(\omega)$ the refraction index of the material, $\epsilon_{0}$ the permittivity of vacuum,
$c$ the speed of light, and $\text{Re}\sigma(\omega)$ the absorption part of the optical
conductivity, which only depends on the size of the nodal line, i.e., $\text{Re}\sigma(\omega)=\frac{e^{2}}{h}\frac{\pi}{8}\sqrt{m/B}$.
Due to the cubic symmetry of the original Hamiltonian, $\delta$ is also
assumed to be isotropic.

In the following, we also take Cu$_{3}$NPd, a material candidate of NLSM with type-I
crossing,  as a concrete example to estimate the Hall conductivities and the Hall
voltages. Cu$_{3}$NPd crystallizes in the cubic perovskite structure with lattice
constant $a=3.85$ {\AA}\cite{Weber1996Cu3NPd}. First principle calculations found that the nodal-line size (diameter)
is about $0.4\pi/a$, i.e., $\sqrt{m/B}\sim 0.2\pi/a$\cite{Kim2015,Yu2015}. For $\theta=0$,
the angle that creates double-Weyl
points,  we have
\begin{eqnarray}
|\sigma_{xy}|&=&2\frac{e^{2}}{h}\frac{\sqrt{\tilde{m}/B}}{\pi}\approx 4\times 10^{4}\Omega^{-1}m^{-1},\nonumber\\
\sigma_{yz}&=&0.\label{conductivity}
\end{eqnarray}
We do not find any experimental result of the refractive
index $n(\omega)$ of Cu$_{3}$NPd, but for its parent material Cu$_{3}$N,
the refractive index $n(\omega)$ is about $3$ in the visible light regime\cite{Odeh2008Cu3NPd}, and we take this value
to estimate the penetration depth. Thus,
\begin{eqnarray}
\delta(\omega)&\approx&\frac{n(\omega)\epsilon_{0}c}{\pi^{2}e^{2}/40ha},\nonumber\\
&\approx&\frac{3\times8.85\times10^{-12}\times 3\times10^{8}\times40\times6.63\times10^{-34}}{3.14^{2}\times(1.6\times10^{-19})^{2}}a,\nonumber\\
&\approx&834a\approx320 nm.\label{depth}
\end{eqnarray}
For the dc conductivity of Cu$_{3}$NPd,  the room temperature value\cite{Ji2013Cu3NPd} is about
$1\times10^{5}\Omega^{-1}m^{-1}$.
The value in the zero temperature limit is expected to be larger, and we assume $\sigma_{yy}=5\times10^{5}\Omega^{-1}m^{-1}$ as an estimation.
We take $l_{x}=l_{y}=100$ $\mu$m, $d=500$ nm,
and $I_{y}=100$ mA, then a combination of Eq.(\ref{voltage}), Eq.(\ref{conductivity})
and Eq.(\ref{depth}) gives $V_{x}\approx20$ mV, which is well within
the capacity of current experiments.

\end{document}